\documentclass[twocolumn,jpsj2,showpacs,preprintnumbers,amsmath,amssymb]{jpsj2}
\def\runtitle{Million-atom molecular dynamics simulation
by order-N electronic structure theory and 
parallel computation}
\def\runauthor{Masaaki Geshi, Takeo Hoshi and Takeo Fujiwara}

\title{%
Million-atom molecular dynamics simulation
by order-N electronic structure theory and 
parallel computation}

\author{%
Masaaki Geshi,
 Takeo Hoshi and Takeo Fujiwara
}

\inst{%
Department of Applied Physics, University of Tokyo,\\
Hongo 7-3-1, Bunkyo-ku, Tokyo 113-8656\\
}

\recdate{\today}

\abst{%
Parallelism Of Tight-Binding Molecular Dynamics  
Simulations Is Presented By Means Of The Order-N Electronic 
Structure Theory With The Wannier States,
Recently Developed (J. Phys. Soc. Jpn. {\bf 69},3773 (2000)).
An Application Is Tested For Silicon Nanocrystals Of More Than 
Millions  Atoms With The Transferable Tight-Binding Hamiltonian. 
The Efficiency Of Parallelism Is Perfect, 98.8\%,
And The Method Is The Most Suitable To  Parallel Computation. 
The Elapse Time For A System Of $2\times 10^6$ Atoms Is 
3.0 Minutes By A Computer System Of 64 Processors Of Sgi Origin 3800.
The Calculated Results Are In Good Agreement With The Results 
Of The Exact Diagonalization,  
With An Error Of 2\% For The Lattice Constant 
And Errors Less Than 10\% For Elastic Constants. 
}

\kword{%
order-N method, parallel computation, tight-binding Hamiltonian, 
silicon nanocrystal}

\begin{document}
\sloppy
\maketitle
\section{Introduction}
Accurate large-scale atomistic simulations are very important 
to investigate and to predict various properties of materials.  
For this purpose, the first principle electronic structure theories 
have been extended  to  calculations of the total energy and forces 
and the first principle molecular dynamics (MD) simulation or the 
Car-Parrinello method~\cite{CP} are now used quite widely   
in the condensed matter physics. 
However, the systems for the first principle MD simulations 
are practically limited to much smaller size, at most, 
of hundreds atoms and much shorter time period of few tens 
pico-seconds.
The other extreme is the classical MD simulations 
with short-range interatomic potentials which are applied 
to systems of millions or ten millions atoms 
with time period of a few hundreds pico-seconds.~\cite{CL,Vash} 
Classical MD simulations are very useful to investigate 
nanoscale systems when  accurate interatomic potentials can be used. 
Even so,  applicability of classical MD simulations 
is limited to phenomena in which electronic process does not
play an essential role.

Modern material technology is deeply involved  
in   electronic processes. 
Then intense attention has been paid to the order-N method 
for the electronic structure calculations, 
whose computational cost increases in 
{\it linearly} proportion to the number of 
electrons.~\cite{Goedecker99a,Ordejon98a}

Novel order-N method is being developed on the 
basis of the Wannier states.~\cite{hoshi2000,hoshi2001}
The Wannier states is formally defined with the unitary transformation 
of the occupied eigen states. 
Once we get true Wannier states $|\psi_j \rangle$, 
the density matrix can be defined as 
\begin{eqnarray}
\hat{\rho} = \sum_j^{occ} | \psi_j \rangle \langle \psi_j|  \ .
\label{rho-exact}
\end{eqnarray}
The expectation value of any physical quantity $X$ 
can be obtained with the density matrix or the 
Wannier states  as 
\begin{eqnarray}
\langle {\hat X} \rangle = {\rm Tr}[{\hat \rho} {\hat X}]
 =  \sum_j^{occ} \langle  \psi_j |{\hat X} |\psi_j \rangle .
\end{eqnarray}
If we put the localization constraint to construct  approximate 
Wannier states with a loss of certain amount of accuracy, then  
we can formulate the order-N method and 
reduce  computational cost extremely. 
Using this order-N method, a system of $1.4 \times 10^6$ atoms was 
calculated by a single CPU standard workstation \cite{hoshi2003}.

In the present paper we do  parallel computation of the perturbation
procedure of the order-N method, 
we call the perturbative order-N method. 
Silicon nanocrystals are calculated up to a system of 2,097,152 
atoms, using SGI Origin 3800 system, and 
the efficiency of parallelism is analyzed. 
To test an accuracy and applicability for calculation of physical 
quantities, 
the lattice constant and elastic constants are 
calculated using cluster systems of up to 1,423,909 atoms.
The usefulness and the limit of the perturbative order-N method 
will be discussed in detail. 

\section{Theoretical backgrounds}
\subsection{The Wannier states}

The Wannier states centered on the $j$-bond can be expressed  as 
\begin{equation}
|\psi_j \rangle =
  C_j^{(0)}|b_j\rangle + \sum_{i(\ne j)} C_j^{(i)} |a_i\rangle \ , 
\label{eq4a}
\end{equation}
where $C_j^{(0)}$ is the  mixing coefficient
of the central bonding orbital $|b_j\rangle$ 
and  $C_j^{(i)}$ is that of the anti-bonding orbital $|a_i\rangle$ 
on the neighboring $i$-bond.~\cite{hoshi2000,hoshi2001}
The mixing of the bonding orbitals on the neighboring bonds 
are negligibly small  due to the orthogonality and the completeness,  
because they contribute to other Wannier states.

For diamond structure crystals, 
we adopt the transferable Hamiltonian 
$\hat H$ of Kwon {\it et al.}~\cite{Kwon} The Hamiltonian includes  
the tight-binding interactions and  
the short-range repulsive interactions between ion cores.  
We truncate  the hopping interactions between the first and the second 
neighbor distances. 
If we denote sp$^3$ hybridized orbitals $|h_i\rangle$,  
the bonding orbital $|b_j \rangle $ and the anti-bonding orbital 
$|a_j \rangle $ are linear combinations of the two hybridized orbitals 
$\bigl( |h_i \rangle \pm  |h_{i^\prime} \rangle\bigr)/\sqrt{2}$.

In the case of silicon crystals, 
the exact results for a system of 512 atoms are  $|C_j^{(0)}|^2 =0.938$ and   
$\sum_i|C_j^{(i)}|^2$ up to the second bond-steps is 0.995.
On the other hand, by the first-order perturbation theory 
the coefficients can be given by an equation  
\begin{equation}
\frac{C^{(\nu (i))}}{C^{(0)}} =
  \frac{\langle a^{(\nu (i))}|\hat{H}|b_k \rangle}
     {\varepsilon_b - \varepsilon_a} \ ,
\label{eq4b}
\end{equation}
and this gives $|C_j^{(0)}|^2 =0.934$.~\cite{hoshi2000,hoshi2001} 
Note that, the first-order perturbation theory gives the value 1 for 
$\sum_i|C_j^{(i)}|^2$ up to the second bond-steps.

\subsection{Perturbative order-N method}

The total energy in the tight-binding formalism is given as 
\begin{eqnarray}
E_{\rm tot}=E_{\rm bs}+E_{\rm rep} ,
\end{eqnarray}
where $E_{\rm bs}$ is the  band structure (BS) energy  
and $E_{\rm rep}$ is the  repulsive energy. 
On the basis of the Wannier states $|\psi_j \rangle$,
the BS energy and its contribution to forces on 
the $I$ atom (site ${\bf R}_I$) are written, 
with the tight-binding Hamiltonian ${\hat H}$,  as~\cite{hoshi2000} 
\begin{eqnarray}
E_{\rm bs} \equiv  {\rm Tr}[\hat{\rho} \hat{H}]  
= \sum_j^{occ} \langle \psi_j|\hat{H}|\psi_j \rangle  
\label{eq2p}
\end{eqnarray}
and 
\begin{eqnarray}
{\bf F}_I^{\rm bs} 
\equiv  {\rm Tr}[\hat{\rho} \frac{\partial \hat{H}}{\partial {\bf R}_I}]  
 = - \sum_j^{occ} 
 \langle \psi_j|\frac{\partial \hat{H}}{\partial {\bf R}_I}| \psi_j \rangle \ 
 .           
\label{eq3p}
\end{eqnarray}
We will calculate the  Wannier states 
by using the perturbative treatment Eqs. (\ref{eq4a}) and (\ref{eq4b})  
and the  density matrix should be given in the same equation 
as Eq. (\ref{rho-exact}) with calculated Wannier states 
$|\psi_j \rangle$. 
The computational cost of the procedure is linearly scales by the 
number of electrons $N$,~\cite{hoshi2000,hoshi2001} and
this procedure we call the perturbative order-N method.

When we use the variational procedure to obtain the Wannier states, 
the physical quantities should be calculated 
in a way consistent with the calculation of 
the Wannier states.~\cite{hoshi2000,hoshi2001}    
Therefore, the density matrix in the above Eqs. (\ref{eq2p}) 
and ({\ref{eq3p}) should be replaced by the optimal one 
${\tilde \rho} = 2{\hat \rho}-{\hat \rho}^2$, and 
this procedure we call the variational order-N method.

\subsection{Linearly scaling property of perturbative order-N method}

The Wannier states and the matrix elements 
of the Hamiltonian are given on the basis of the atomic orbitals 
$|\phi_{I \alpha}\rangle$. 
Then we can estimate the computational cost in the following way. 
Firstly  the above physical quantities can be rewritten, 
by using the matrix elements of the Hamiltonian 
and the density matrix represented by the atomic orbitals, as
\begin{eqnarray}
E_{\rm bs} 
&=& \sum_I^{N_{\rm atom}}\sum_{\Delta}^{N_{\rm loc}}
  \sum_{\alpha}^{N_{\nu}}\sum_{\beta}^{N_{\nu}} 
     \rho_{I \alpha (I+ \Delta) \beta} \nonumber \\
& & \mbox{\hspace*{1.0cm}} \times 
\langle \phi_{I \alpha} | \hat{H} | 
     \phi_{(I+\Delta) \beta} \rangle, 
\label{eq2}
\end{eqnarray}
and 
\begin{eqnarray}
{\bf F}_I^{\rm bs} 
 &=& -\sum_{J}^{N_{\rm atom}} 
      \sum_{\Delta}^{N_{\rm loc}}
    \sum_{\alpha}^{N_{\nu}} \sum_{\beta}^{N_{\nu}}
    \rho_{J \alpha (J+ \Delta) \beta} \nonumber \\
 & & \mbox{\hspace*{1.0cm}}  \times\langle \phi_{J \alpha}
       |\frac{\partial\hat{H}}{\partial {\bf R}_I}|
       \phi_{(J+ \Delta) \beta}\rangle   \ . 
\label{eq3}
\end{eqnarray}
The numbers $N_{\rm atom}$, and $N_{\nu}$ are those  of atoms and 
atomic orbitals per atom, respectively.
The number $N_{\rm loc}$ is that of interacting atoms 
in the local region around the central atom. 
The local region is defined, outside which the matrix elements of 
the tight-binding Hamiltonian  vanish.
In the diamond structure, $N_{\rm loc} =17$ 
including the central, the first and the second neighbor atoms.

From Eqs.~(\ref{eq2}) and (\ref{eq3}), 
the total computation time of the matrix elements are scaled by a factor 
$N_{\nu} \times N_{\nu}\times N_{\rm loc} \times N_{\rm atom}$, 
where the factor  $N_{\nu} \times N_{\nu}$ is due to the cost 
for  the quantum mechanical calculation.
In the sp$^3$ minimal basis set, $N_{\nu}$ is four.
Therefore, we can calculate each Wannier state by a local
procedure and the total computation cost is proportional 
to the number $N_{\rm atom}$. 
The procedure is then the perfect order-N method.

Non-negligible  amount of computation time is consumed 
in the calculation of the repulsive energy and forces. 
The part of the listing  of the neighboring atoms 
is also important and its cost is not negligible  for large systems.
The above two parts are the same as in standard classical 
MD simulations and  
the  computation time of these two parts are scaled by a factor
$N_{\rm loc} \times N_{\rm atom}$. 
The computation time of the BS energy and forces costs, at least, 
$N_{\nu} \times N_{\nu}$ times more than those of 
the calculation of repulsive interactions 
and the listing of the neighboring atoms.

\subsection{Allotment of Wannier states to processors, 
memory size and communication of data}

Since the perturbative treatment is  completely independent 
among the Wannier states, we can parallelize the computation 
with respect to several groups of the Wannier states.  
When we use $N_{\rm CPU}$ processors, each processor participates 
in the calculation of about $N/N_{\rm CPU}$  states among the 
total $N$ Wannier states.
For example, the calculation of \{
$ \langle \psi_j|\hat{H}|\psi_j \rangle $ and 
$ \langle \psi_j|\frac{\partial \hat{H}}{\partial {\bf R}_I}| \psi_j \rangle $
\}$_{j=j_{n-1}+1, \cdots ,j_n}$ \ ,  ($j_{n}-j_{n-1} \simeq N/N_{\rm CPU})$ 
is allotted to the $n$-th processor.

The matrix elements are not stored on memory because 
they require totally a large CPU memory. 
For example, the total memory size for the matrix elements 
$ \bigl\{ \langle \phi_{I \alpha} | \hat{H} | 
  \phi_{(I+\Delta) \beta} \rangle \bigr\}$
can be estimated to be 
$8~({\rm B}) \times 4^2 \times 17 \times 10^6 = 2.2~({\rm GB})$ for 
a system of $10^6$ atoms. 
Therefore, we  calculate the matrix elements when 
they are required and do not store them.

In the case of $10^6$ atoms, 
the memory size for  all atomic positions is 
$8~({\rm B})\times 3 \times 10^6 =24~({\rm MB})$,  
that for the listing of the neighboring atoms is 
$4~({\rm B})\times 16 \times 10^6 =64~({\rm MB})$, 
and  that for the force is 
$8~({\rm B})\times 3 \times 10^6 =24~({\rm MB})$. 
Then the total memory size for each processor is not large.
In the present calculation,   
all processors have the same data of all atomic positions and  
no data are communicated  among processors during 
the calculation of the Wannier states and contributions 
to the BS energy and forces.

After  calculating contributions to the BS energy and forces 
in individual processor, 
we should sum up these elements. 
This procedure is accomplished by MPI\_ALLREDUCE command in the 
Message Passing Interface (MPI) in the present calculation.
This is wasteful procedure with respect to the communication of the data 
because a large number of communicated data is not necessary, 
that is, null.
Notwithstanding, the communication time would  be relatively 
cheap expenses in the present work 
because there exist very heavy computations 
with respect to  quantum mechanical freedoms.

\begin{figure}[t]
\begin{center}
\includegraphics[width=8.0cm]{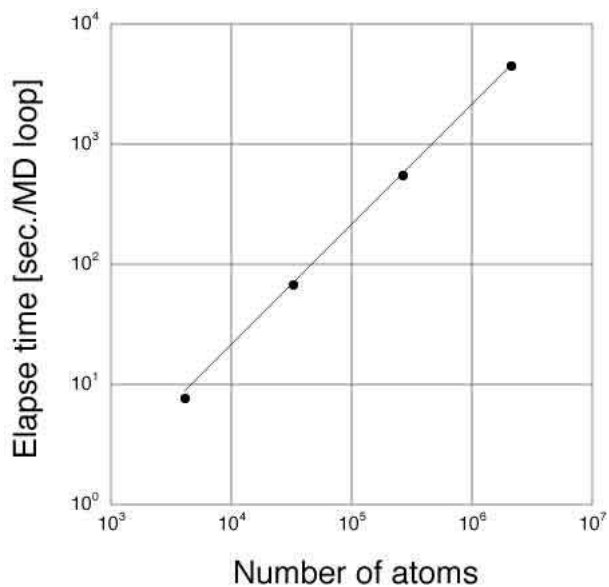}
\end{center}
\caption{The elapse time of one MD loop by using one processor 
for the systems of 
4,096,  32,768,  262,144  and 2,097,152 atoms of Si crystals. 
}
\label{fig:1}
\end{figure}

\section{Results and discussions}

\subsection{Elapse time of one processor}

We calculated four different systems of 
4,096,  32,768,  262,144  and 2,097,152 atoms 
of Si crystals with the periodic boundary condition. 
Figure \ref{fig:1} shows the elapse time of one MD loop 
and one can see the almost perfect linearly scaling property.
The elapse times for the calculation of the BS energy and forces 
are 97.44  \%, 97.65\%, 97.63\% and 97.64\% 
of the total elapse times in respective systems, 
which are larger than a factor 
$16/(16 + 1) = 0.941$, 
estimated simply  in the case of $N_{\nu} \times N_{\nu}=16$.
Note that the small difference between two numbers 
0.976 and 0.94 is very serious for the speed-up ratio 
discussed later.  
From these results, we can conclude that the tight-binding calculation 
is heavier by a factor 2.5 $(=\frac{0.976}{1-0.976}\times \frac{17-16}{16})$ 
than our naive estimation. 
This is very crucial difference in actual simulations,
and quantum mechanical calculations are much heavier 
than the classical simulations 
even in the tight-binding calculation. 
One should parallelize firstly the part of the calculation of 
the BS energy and forces. 
In practice, we parallelize also the part of the calculation 
of the repulsive interaction energy.  
The parallelizable fraction $P$, the fraction of the elapse time 
of strictly parallelizable part among  the whole elapse time  
in the case of one processor,  reaches to 0.988. 
This high value of $P$ would be impossible without 
the generic property of the linear scaling of the present method.   
Therefore, the perturbative order-N method is one of the most 
suitable procedures to the parallel computation.

\subsection{Speed-up ratio by parallelism}

When we parallelize the computational program 
by using $N_{\rm CPU}$ processors, the speed-up ratio 
$ \alpha_p $ is defined as the ratio
of the elapse time $t_{N_{\rm CPU}}$ of $N_{\rm CPU}$ processors and 
that $t_1$ of one processor as
\begin{eqnarray}
\alpha_p \equiv \frac{t_1}{t_{N_{\rm CPU}}} \ .
\label{sd_ratio}
\end{eqnarray}
Let us assume that  
we can parallelize the part of the fraction $P$ perfectly. 
In other words, we assume that the elapse time of this part can be 
reduced by a factor ${1}/{N_{\rm CPU}}$. 
In such strictly parallelized case, 
the total elapse time can be minimized  with 
the maximal speed-up ratio~\cite{Amdahl} 
\begin{eqnarray}
  (\alpha_p)_{\rm max}
   =     t_1 \times \Bigl(\frac{1}{t_{N_{\rm CPU}}}\Bigr)_{\rm max}
 \equiv  \frac{1}{(1-P)+\frac{\displaystyle P}{\displaystyle N_{\rm CPU}}} \ .
\end{eqnarray}
The high speed-up ratio is possible  only for 
the high value of the parallelizable fraction $P$.

\begin{figure}[t]
\begin{center}
\includegraphics[width=8.0cm]{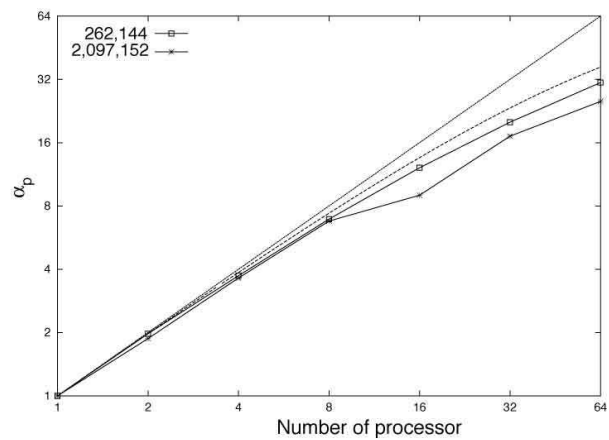}
\end{center}
\caption{
 The speed-up ratio  $\alpha_{p}$ 
for systems of 262,144  and  2,097,152 atoms  
as a function of the number of processors. 
That for a system of  32,768 atoms locates 
slightly above that of 262,144 atoms. 
The  dashed lines are the maximum speed-up ratio 
$(\alpha_{p})_{\rm max}$ of $P=0.988$ and 
dotted lines are just $(\alpha_p)_{\rm max} = N_{\rm CPU}$ 
corresponding to $P=1$.
The observed deviation of the line of 2,097,152 atoms 
at 16 processors is due to the consumption of the elapse time 
in the data communication, 
presumably because of an ill balancing of the 
data size and the number of processors. 
}
\label{fig:2}
\end{figure}

\begin{center}
\begin{figure*}[t]
\includegraphics[width=17.0cm]{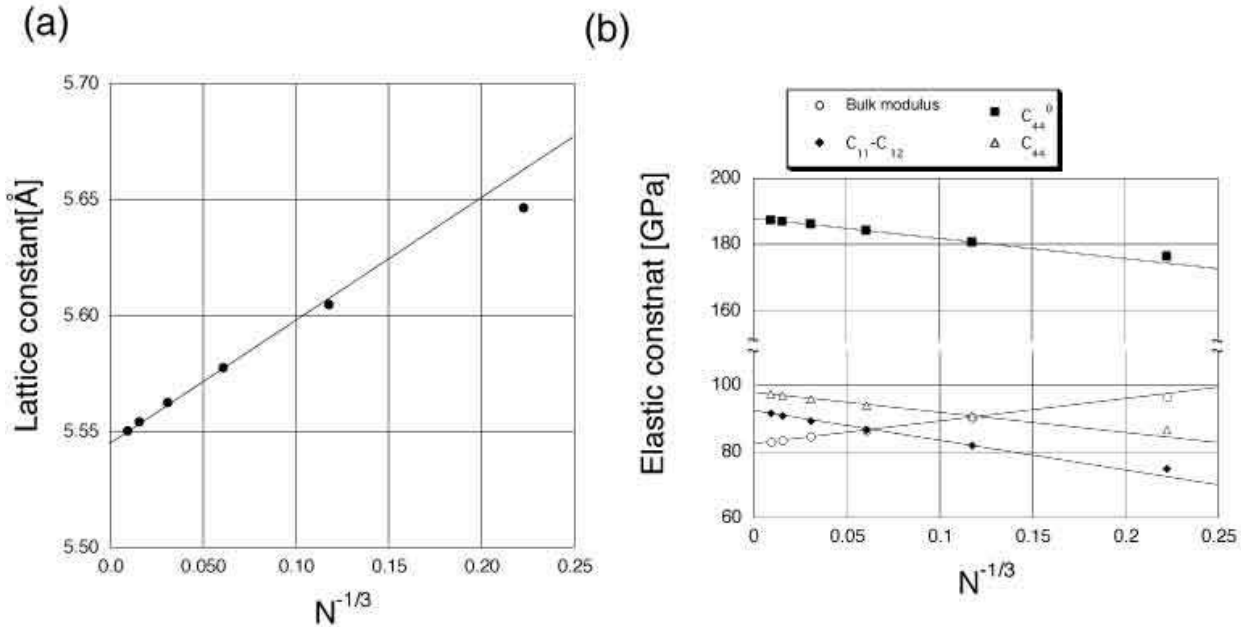}
\vspace{-1.0cm}
\caption{The lattice constant (a) and elastic constants (b) 
as a function of $N^{-1/3}$, obtained by the perturbative method.
}
\label{fig:4}
\end{figure*}
\end{center}

\begin{table*}
\begin{center}
\begin{tabular}{@{\hspace{\tabcolsep}\extracolsep{\fill}}rrrrr} 
\cline{1-5}
    &  Present work &Diag. ~\cite{Kwon}& LDA ~\cite{NM} &  Exp.\\ \cline{1-5}
Lattice constant[\AA]& 5.546 (2.19\%) & 5.427  &  5.431 & 5.429  ~\cite{Donohue}\\
Bulk modulus    [GPa]& 82.3 (6.05\%) & 87.6   & 93.0   &  97.8  ~\cite{McSkimin} \\
$C_{11}-C_{12}$ [GPa]& 92.3 (1.70\%) & 93.9   & 98.0   & 101.2  ~\cite{McSkimin} \\
$C_{44}^0 $     [GPa]& 188.3(5.14\%) & 198.5  &111.0   &        \\
$C_{44}$        [GPa]& 97.9 (10.0\%) & 89.0   & 85.0   &  79.6  ~\cite{McSkimin} \\
\cline{1-5}
\end{tabular}
\end{center}
\caption{Lattice constant and elastic constants extrapolated 
to $N \rightarrow \infty$ systems are compared with those by 
the diagonalization method (Diag.), 
the first-principles calculation within the Local Density 
Approximation (LDA), and experiments (Exp.). 
The values in parentheses are 
errors against those with the diagonalization method.
The elastic constant $C_{44}^0$ is that in a system where the internal
displacement is not relaxed.}
\label{table1}
\end{table*}

 Figure \ref{fig:2} shows the  observed speed-up ratio 
$\alpha_{p}$ by using MPI and 
the maximum one $(\alpha_{p})_{\rm max}$ with  $P=0.988$
as a function of the number of processors. 
Compared with the minimum  elapse time   $t_1/({\alpha_p})_{\rm max}$, 
the present computation consumes   2~\%  more for 32,768 atoms (8 processors), 
18~\% more for 262,144 atoms (64 processors), 
and  45~\% more for 2,097,152 atoms  (64 processors), respectively.
In the case of 64 processors for 2,097,152 atoms, the communication 
cost is about 30~\% among whole elapse time. 
The cost of data communication increases with the number of 
the processors. 
The above analysis indicates that, 
when one uses a computer system of more than 200 processors for 2,097,152 atoms,
the  data communication process would consume  
the majority of the elapse time.

For a system of $2,097,152$ atoms, the elapse time of one MD loop by  
one processor is 4565~s (76.1~min). 
When the time interval of one MD step corresponds to 3 femto seconds 
in physical systems, 
a one pico second simulation needs 422~hours (17.6~days) of CPU time 
by one processor.  
Such simulation becomes feasible when we use 128 processors.  

\subsection{Calculations of elastic constants}

The lattice constant and elastic constants are 
calculated in systems of several sizes of clusters of Si crystal 
of up to  1,423,909 atoms. 
The  boundary condition is such that  
hybridized orbitals of the ideal sp$^3$ type are fixed 
on the surface atoms but  surface atoms can move.  
Since the elastic constants are the linear response to small 
distortions, they may be expected to be reproduced by the 
first order perturbation calculations.

Figure \ref{fig:4} shows the calculated results  with increasing
the number of atoms $N$.
The deviation can be scaled by $N^{-1/3}$ 
because it is the effects of the  surface. 
The values extrapolated to $N \rightarrow \infty$ are summarized in 
Table \ref{table1}.
The present results agree with those of the exact diagonalization
method ~\cite{Kwon} within less than 10\% error.
The errors except for $C_{44}$ are not more than the difference 
between results by the present tight-binding Hamiltonian (Diag.)  
and the LDA.
The deviation of the shear modulus $C_{44}$ is larger than those of 
the bulk modulus or $C_{11}-C_{12}$, 
since  $C_{44}$ is inherently complicated 
due to the rehybridization and the internal distortion~\cite{Harrison,NM} and 
this phenomena cannot be described very accurately  
by the first order perturbation 
of the  Wannier states of fixed sp$^3$ hybrids.
The discrepancy between the results with the order-N method and 
those with the diagonalization method originates from the localization 
constraint for constructing the Wannier states and the perturbation 
treatment in Eqs.~(\ref{eq4a}) and (\ref{eq4b}).  
These restrictions are controllable and  
the  discrepancy is not serious  between  the results by 
the present order-N method and the diagonalization method. 
This is a typical example of account balancing between 
the accuracy and the computational cost in the order-N method.

Much larger error is found  in the value of $C_{44}^0$ 
of the tight-binding calculation itself,  
compared with that of  the LDA calculation, 
which is a limitation of the present tight-binding Hamiltonian. 
One of the important works in future is the construction 
of more accurate tight-binding Hamiltonian from the first principle 
electronic structure calculations. 
Even if such sophisticated Hamiltonian is much complicated, 
it does not cause any essential difficulty in the calculation 
by the present order-N method, though it may increase the CPU times.

\section{Conclusion}

In a summary, we demonstrated the efficiency of parallelism of the 
perturbative order-N method in the large-scale tight-binding 
MD simulations. 
The method was shown to be the most profitable procedure for 
the parallel computation. 
The communication time is, even in the present case,
much less than the time of the quantum mechanical calculations.

The perturbative order-N method may be hardly applied to 
systems with large distortion of lattices or bond breaking 
because the deviation from the unperturbed states 
becomes very large and, in the bond breaking process, 
the charge transfer and  re-bonding 
are essentially important. 
In such cases we should combine the perturbative order-N method
with other methods for constructing basis states. 
The variational  method can be associated 
with the perturbative order-N method and we can compose a hybrid order-N 
method. 
The hybrid order-N method can give electronic structures in the whole
system and, more importantly, there is no discontinuous 
boundary in the connected region. 
The hybrid scheme of the perturbative and variational order N-methods 
has been already applied to the fracture 
propagation in Si nanocrystals with 1.4 $\times 10^6$  atoms 
without parallelism~\cite{hoshi2003}.
The parallelized hybrid order-N method is 
a very essential method to pursue tight-binding MD simulations  
for systems of millions atoms and 
the perturbative order-N method 
is the  key technique in order to enlarge the size of 
the whole systems.

\section*{Acknowledgments}
We are very grateful for useful discussion about general 
techniques of parallelism by MPI with F. Shimizu and H. Kimizuka.
Computation has been done at the Center for Promotion of Computational 
Science and Engineering (CCSE) of Japan Atomic Energy Research 
Institute (JAERI) and also partially carried out by use of the 
facilities of the Supercomputer Center, Institute for Solid 
State Physics, University of Tokyo.  
This work is financially supported by Grant-in-Aid from 
the Ministry of Education, Culture, Sports, Science and 
Technology 
and also by ``Research and Development for
Applying advanced Computational Science and Technology'' 
of Japan Science and Technology Corporation.


\end{document}